\documentclass[aps,prd,12pt,superscriptaddress,
amsfonts,amssymb,amsmath,byrevtex,showpacs,nofootinbib]{revtex4-1}
\pdfoutput=1
\usepackage{graphicx}
\usepackage{amssymb,amsmath,amsfonts,palatino,amsthm}
\usepackage{amssymb}
\usepackage{epstopdf}
\usepackage{color}

\begin{document}

\title{Quantum Gravity on a Quantum Chip\footnote{Essay written for the Gravity Research 
Foundation 2018 Awards for Essays on Gravitation. \\ Submitted on 28 March 2018.}}

\author{Jakub Mielczarek\footnote{jakub.mielczarek@uj.edu.pl} \\
CPT, Aix-Marseille Universit\'e, Universit\'e de Toulon, CNRS, F-13288 Marseille, France \\
Institute of Physics, Jagiellonian University, {\L}ojasiewicza 11, 30-348 Cracow, Poland}

\begin{abstract}
This essay aims at emphasizing the potential of a synergy between quantum gravity 
and the quantum computing technologies. Such a combination would be beneficial for both 
understanding the Planck scale physics and the stimulation of development of the quantum 
technologies. This is especially important in the present early days of commercial quantum 
computers, when challenges originating from the basic research may catalyze the technological 
progress.  Our attention is focused on simulations of a Planck scale system with the use of 
existing adiabatic quantum computers. Current possibilities, technological challenges and 
prospects for the future are outlined. 
\end{abstract}

\maketitle 

\section{Introduction}

Theoretical physics is a source of challenges that catalyze the development of new technologies. 
This factor is especially crucial in the infancy of an emerging technology, when the commercial 
market is not yet ready to play the role of a driving force. On the other hand, advanced 
technologies make the progress in theoretical physics possible. Therefore, theorists should also look 
for new technological possibilities that may allow to deepen our basic knowledge. 

There are many profound examples of mutual benefits coming from the interplay 
between theoretical physics and the advanced technologies. One of the most 
tangible is the emergence of World Wide Web, as a result of the need to improve 
exchange of information between particle physicists working on experiments at CERN. 
The theoretical physics provided the reason to search for new particles and phenomena, 
verification of which required the development of novel technological solutions. The exploration of virgin 
areas in high energy physics, like the successfully accomplished hunt for the Higgs particle, 
incessantly pushes the technological boundaries forward. The obtained technological 
solutions not only allow to deepen our knowledge in the basic research but they simultaneously 
diffuse to commercial applications, changing our everyday life. The links and the impact are 
not always as clear as in the case of Internet but the examples are around. For instance, 
the technology of superconducting magnets, which is crucial for particle accelerators, 
plays an important role in the medical imaging devices such as MRI. 

Worth stressing here is that such a synergic progress may not always be possible. In particular, 
if the realm of some theoretical considerations is too far from the reach of the existing technology, 
any attempt of driving the technological progress by the basic research is doomed to fail. For instance, 
supposing that quantum mechanics was miraculously discovered in Middle Ages, the 
technology of the time was too primitive to gain anything from such a theoretical insight. 

The latter situation is, however, quite unlikely because levels of development of the basic research 
and technology are mutually related and there are no such extreme divergences between 
the two areas of human activity being observed. The technology is rather tracking where the 
basic research is and contributes there, as well as vice versa. In consequence, both disciplines 
are typically following in the common direction, even if the resulting benefits may be of 
completely different type. 

The aim of this essay is to stress the importance of a new area where the symbiosis between 
theoretical physics and new technologies may lead to significant progress in both disciplines. 
The discussed area is located at the interface of the Planck scale physics and quantum computing 
technologies. 

\section{Quantum simulations of the Planck scale physics}

The Planck scale physics may seem to be too far from our reach to drive any technological progress. 
The associated energy scales are around fifteen orders of magnitude above the maximum 
energies achieved with the use of the current accelerator technologies. Possibilities for directly 
probing the realm of quantum gravity are still very distant (see e.g. \cite{Hossenfelder:2010zj}). 

However, one piece of technology is actually being broadly used to study quantum gravity -- the 
computers. The classical data processing machines provide the computational power 
for both symbolic analyses and simulations. We will focus our attention here on the latter. 
Arguably, the biggest success towards simulating quantum gravity has been 
obtained within the so-called Causal Dynamical Triangulations (CDT) \cite{Ambjorn:2012jv}. 
In this approach the quantum system under consideration is investigated in the 
path integral formalism of quantum mechanics. First, a discretization of the continuous 
gravitational field is performed in order to reduce the number of degrees of freedom. 
The second necessary step is to perform the Wick rotation, so that the original quantum system is 
converted into a statistical system, which can be simulated with the use of the Monte Carlo 
methods. 

The CDT approach is an example of simulating quantum systems on a classical computer. 
However, would not it be better to use quantum computers to study the quantum systems? 

The answer seems to be affirmative. The only caveat is that the accessible quantum 
computational power is still very limited. Nevertheless, recent years have brought the enormous 
progress in the quantum computing technologies. It is mostly thanks to the development of 
superconducting quantum circuits \cite{You}, which provide scalable quantum information 
processing devices. 
 
Simulations of quantum systems on the quantum computers are very different from those 
performed on the classical machines. Actually, quantum computers perform real experiments 
on certain quantum systems, which allows for the so-called \emph{exact simulations} 
\cite{Feynman:1981tf}. It means that a quantum system under consideration is mapped into 
another quantum system, which is engineered with the use of the architecture of a quantum chip. 

From the quantum mechanical perspective, different physical systems may have the same 
mathematical description. For instance, eigenvalues of a quantum harmonic oscillator 
are the same irrespectively of the underlying physical realization (e.g. the harmonic approximation of a 
diatomic molecule, the monochromatic quantum light, a quantum LC circuit, etc.). The same 
concerns quantum simulations of the Planck scale physics. 

Degrees of freedom of the Planck scale system can be mapped into an 
equivalent set of quantum degrees of freedom, which will \emph{imitate} the original system 
under interest. In this way, the previously experimentally inaccessible degrees of freedom 
can be exactly reproduced by some other degrees of freedom, which can actually be controlled 
and measurements on which can be performed. 

\begin{figure}[ht!]
\centering
\includegraphics[width=12cm,angle=0]{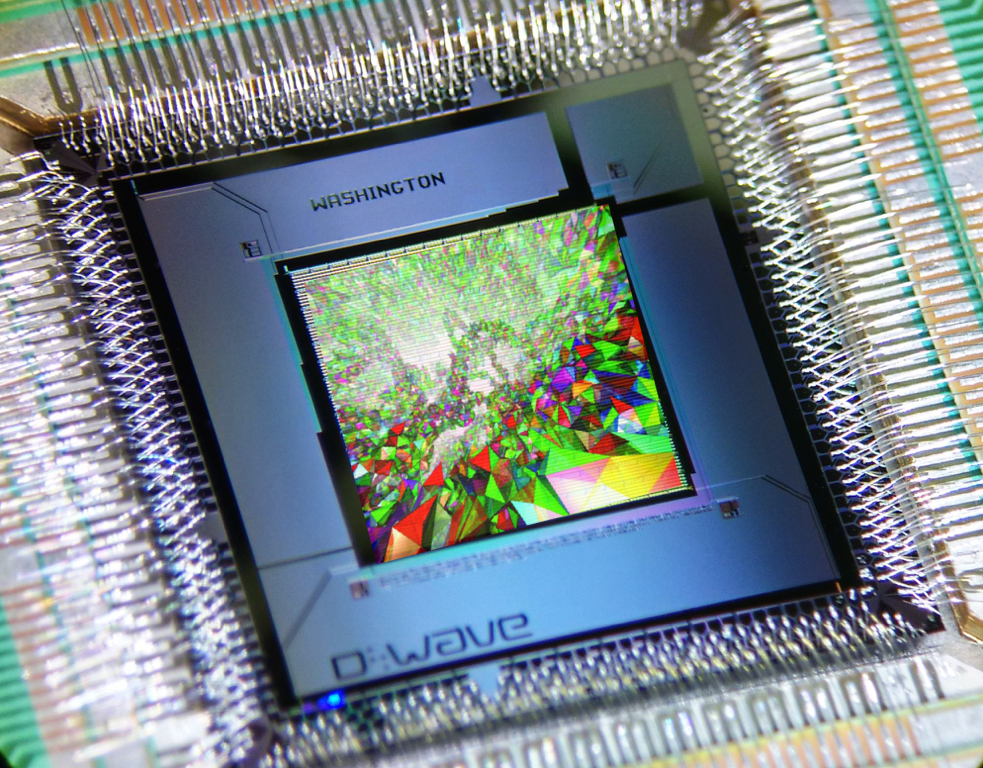}
\caption{A collage of Loop Quantum Gravity being simulated 
on a D-Wave superconducting quantum chip. A pictorial representation 
of the Planck scale physics \cite{LQG} and a picture of the D-Wave quantum 
chip \cite{DWAVE} have been used.}
\label{QuantumChip}
\end{figure}

In particular, the commercially available adiabatic quantum computers \cite{McGeoch} allow to map quantum 
gravitational degrees of freedom onto superconducting qubits arranged at a quantum chip. 
This idea is expressed in Fig. \ref{QuantumChip}, depicting a D-Wave \cite{DWavewww} quantum 
processor and an ongoing quantum gravitational simulation. As we will discuss below, 
the picture is much more than a futuristic vision and represents a realistic possibility. 

\section{Loop Quantum Gravity}

To be more specific, let us briefly discuss a proposal of simulating Loop 
Quantum Gravity (LQG) \cite{Ashtekar:2004eh,CRFV} on an adiabatic 
quantum computer, presented in \cite{Mielczarek:2018ttq}. The idea is 
based on the observation that the notion of qubits (two level quantum states) 
naturally emerges for a spin network composed of 4-valent nodes and a 
fundamental ${\rm SU}(2)$ group representation attached to its links \cite{Feller:2015yta,Li:2017gvt}. 
In such a case, the so-called \emph{intertwiner} Hilbert spaces at the nodes are 
two dimensional, since 
\begin{equation}
{\rm dim\ Inv} (H^{1/2}\otimes H^{1/2}\otimes H^{1/2}\otimes H^{1/2})=2\,,
\end{equation}
where ${\rm Inv}$ denotes invariant subspace. 
The resulting spin network is associated with the product Hilbert space 
$\mathcal{H}= \Pi_{i=1}^N \otimes H_{i}^{1/2}$, which represents states of the 
spatial geometry. Here, $N$ denotes the number of vertices. The qubit base states 
$\{|0\rangle,|1\rangle\}$ can be chosen as such that they diagonalize the volume 
operator, i.e. $\hat{V}|1\rangle= +V_0|1\rangle$ and $\hat{V}|0\rangle= -V_0|0\rangle$, 
where $V_0 = \frac{\sqrt{3}}{4} l^3_{\rm Pl}$ is the minimal quantum of volume \cite{CRFV}. 

However, not all qubit configurations are allowed but only those satisfying 
the Hamiltonian constraint $\mathcal{C} \approx 0$. By solving the constraint, the 
physical states $| \Psi_{\rm phys} \rangle$ of the theory can be extracted: 
\begin{equation}
\hat{C} | \Psi_{\rm phys} \rangle  \approx 0\,. 
\end{equation} 
In order to solve the constraint one can equivalently look for the ground states of the 
Master Hamiltonian:
\begin{equation}
\hat{H} \sim \hat{C}^2. 
\end{equation} 
By identifying all degenerate ground states of this Hamiltonian, the physical Hilbert space 
of the original constrained system can be reconstructed. 

The currently available adiabatic quantum computers, such as the D-Wave, allow to extract 
the ground states of the quadratic Hamiltonian 
\begin{equation}
\hat{H} = \sum_{i=1}^N a_i \hat{V}_i+ \sum_{i \neq j}^N b_{ij} \hat{V}_i  \hat{V}_j + const\,, 
\label{IsingModel}
\end{equation} 
which can be associated with a prototype linear constraint 
\begin{equation}
\hat{C} = c\,\hat{\mathbb{I}}+\sum_{i=1}^N c_i \hat{V}_i\,, 
\end{equation} 
such that $a_i = 2 c c_i $ and $b_{ij}=c_i c_j$. However, in the existing adiabatic quantum 
computers not all interactions (couplers) between the qubits are possible but only those 
consistent with the architecture of a given quantum processor \cite{Harris}. Therefore, only certain values 
of $b_{ij}$ and $a_i$ are accessible with the current technology. In particular, those consistent 
with the so-called chimera graph, which represents architecture of the D-Wave quantum chip. 

The adiabatic quantum simulations allow to find physical states of the system and perform 
various measurements on configurations of the quanta of volume. In particular, one can 
measure the correlation functions and the formation of semiclassical domains of the quanta 
of plus volume or minus volume. This enables us to make a progress towards the reconstruction 
of classical spacetime from the Planck scale building blocks. 

\section{Future}

The story of simulating quantum gravity on quantum computers just begins and there are broad 
perspectives for the future, both near and the more distant one. 

The method of simulating LQG deserves further more detailed studies and the actual simulations 
to be performed. The practical realization of quantum simulations is possible to be addressed not 
only in LQG. The quantum annealing algorithm naturally suits the CDT approach, which relies on 
the Monte Carlo optimization process, associated with finding a state of equilibrium. Furthermore, 
quantum fluctuations of black hole horizons are the natural next candidate for being quantum simulated. 

An important issue to stress here is that the two dimensional structure of a quantum chip does 
not necessarily imply that only two dimensional systems can be simulated. What matters is 
the structure of connections between the qubits. Nevertheless, the two dimensional architecture 
imposes certain limits on the number of couplers between qubits. Therefore, the natural path for the 
future development of quantum processors is to go beyond 2D, to the 3D structure of a quantum 
circuit. 

On the other hand, in the case of of gravity, simulations performed on 2D quantum processors 
may turn out to be sufficient to recover the information about 3D quantum gravity. Namely, 
there is increasing theoretical evidence that the spatial 3D geometry may just correspond to the 
quantum entanglement of some system defined at a 2D boundary. The results of AdS/CFT 
correspondence \cite{Maldacena:1997re}, the holographic entanglement entropy \cite{Ryu:2006bv}, 
MERA tensor networks \cite{Swingle:2009bg} and EPR=ER conjecture \cite{Susskind:2017ney} 
contribute to the picture of gravity associated with the entanglement structure of some quantum field 
theory at the boundary. In such a picture, the discussed spin networks of LQG can be perceived 
as the representations of either a state of gravity in the bulk or equivalently the entanglement structure 
of the system at the boundary \cite{Han:2016xmb}. From this viewpoint, simulating quantum 
gravity on a quantum chip should either concern the bulk (as we discussed it so far) or 
degrees of freedom at the boundary. From the second viewpoint, simulations of a quantum 
system at the boundary (e.g. certain conformal field theory) should allow to reconstruct a state of 
quantum geometry in the bulk. 

\section{Summary}

Technology is constantly applying new theoretical achievements to push the limits of engineering 
forward. In turn, theoretical physics should look for new technological possibilities that will allow to 
deepen our basic knowledge. The message of this essay is that there emerges a new direction to 
boost the overall progress thanks to a fusion of quantum gravity and the quantum information 
technologies. What is already becoming possible are simulations of the Planck scale systems with 
the use of existing quantum computers. However, the potential consequences are much broader 
and even difficult to anticipate. Quantum simulations may not only turn out to be a practical tool but 
also allow us to unveil the deeper connections between gravity and the quantum information theory, 
such as the quantum version of the \emph{it from bit} conjecture \cite{Wheeler}. 

\section*{Acknowledgements}

JM is supported by the Sonata Grant DEC-2014/13/D/ST2/01895 of the National Science Centre Poland 
and the Mobilno\'s\'c Plus Grant 1641/MON/V/2017/0 of the Polish Ministry of Science and Higher Education.

\end{document}